\documentstyle[12pt]{article}
\renewcommand{\title}[1]{\bf #1\bigskip\medskip\\} 
\renewcommand{\author}[1]{\large #1\\ \smallskip}
\newcommand{\address}[1]{{\normalsize\it #1\\}\bigskip}
\newcommand{\wt}[6]{#1\mbox{\small
 $\left(\matrix{#5&#4\cr#2&#3\cr}\biggm|\mbox{$#6$}\right)$}}
\newcommand{\Km}[5]{{#1}\biggl(\!\matrix{&#3\vspace*{-0.3cm}\cr\!\!
  #2\hspace*{-0.3cm}\vspace*{-0.2cm}&\cr&#4}\!\!\biggm|\!
  \mbox{$#5$}\biggr)}
\newcommand{\Z}{\mbox{\sf Z\hspace*{-0.45em}Z}}
\newcommand{\cit}[1]{\mbox{\hspace{2pt}}\cite{#1}}
\def\be{\begin{eqnarray}}
\def\ee{\end{eqnarray}} 
\def\no{\nonumber}
\def\T{\mbox{\boldmath { $T$}}}
\def\t{\mbox{\boldmath {$y$}}}
\def\1{\mbox{\boldmath I}}
\begin{document}
\begin{center}
\title{FUSION HIERARCHIES WITH OPEN BOUNDARIES 
         \\  AND EXACTLY SOLVABLE MODELS}
\author{Yu-kui Zhou\footnotetext[1]{\small 
  Email: {\em zhouy@maths.anu.edu.au}}
                  \footnotetext[2]{\small 
  On leave of absence from {\em Institute
     of Modern Physics, Northwest University, 
   Xian 710069,\newline\hspace*{0.5cm} China} } }
\address{Mathematics Department,
        The Australian National University \\
          Canberra, ACT 0200, Australia }
\begin{abstract}
The formulation of integrable models with open boundary conditions
and the functional relations of fused transfer matrices 
are discussed. It is shown that finite-size corrections to the 
transfer matrices and unitarity relations of free energies can be
obtained from the functional relations. Unitarirty relations of 
surface free energies presented in previous papers are also reviewed.
\end{abstract}
\end{center}
\subsection{Introduction} 
It has been well understood that
integrable models with open boundary   conditions
in two-dimensional  statistical mechanics are built on both the
Yang-Baxter equation (YBE)\cit{Yang:67,Baxter:82} and 
boundary YBE\cit{Cherednik}  (reflection equation)
thanks to Sklyanin's work\cit{Sklyanin} in 1988. Since then
there has been a growth of interest in such vertex 
and face integrable models\cite{GhZa:94}$-$\cite{Zhou:95b}.
Integrability of exactly solvable vertex models can be formulated
directly from Sklyanin's study in the six-vertex model
with open boundary conditions. The formulation of  face models 
is then obtainable from the vertex-face correspondence  
(intertwiners\cit{Zhou:95b}). In this short paper, without loss
of generality, we consider Baxter's eight-vertex model\cite{eight}, 
Belavin's $\Z_n$ symmetric vertex models\cit{Belavin}
and their corresponding face models\cit{Baxter:73,JMO:87} 
to formulate integrable models with open boundary
conditions. It is discussed that finite-size corrections to transfer
matrices and free energies can be obtained from the 
functional relations of the transfer matrices. 
Then we review progress in the study of finite-size
corrections to the transfer matrices of the $U_q[sl(2)]$-invariant 
six-vertex model.

In the next section we examine  integrability of the above
models with open boundary conditions. It is clear that  Belavin's $\Z_n$ 
vertex models 
reduce to  Baxter's eight-vertex model for $n=2$. These models 
have recently received considerable 
attention\cit{Baxter:73}$-$\cite{BaRe:90} (for Bethe ansatz of the higher 
rank models see ref.\cite{HYZ:89}). In section~3 we describe functional 
relations and the so-called $y$-system of the fused transfer matrices. 
We discuss how these relations  can be applied to determine both the 
bulk and surface free energies of the models. We also discuss how the 
finite-size corrections to the transfer
matrices are represented by a $y$-system. In section~4
we present  finite-size corrections to the transfer
matrices of an $U_q[sl(2)]$-invariant vertex model\cite{Zhou:95a}. 
\subsection{Exactly solvable models with open boundaries}  
It is well known that the YBE yields Boltzmann weights of  exactly 
solvable models. Let $R_{i,j}^{k,l}(u)$ and $\wt Wa{b}c{d}u$  
be the Boltzmann weights 
of Belavin's vertex models and their face models respectively,
$u$ is the spectral parameter. These Boltzmann weights vanish unless
$i+j=k+l$ (mod $n$) with $i,j,k,l=1,2,\cdots,n$ for the vertex models
and $d-a,b-a,d-c,b-c={\hat 1},{\hat 2},\cdots,{\hat n}$ with 
$a,b,c,d\in w_0+\sum_{\mu=1}^{n}Z_\mu{\hat\mu}$ ($Z_\mu\in\Z$)
for the face models. 
${\hat 1},{\hat 2},\cdots,{\hat n}$ are the elementary vectors of 
the algebra $A^{(1)}_{n-1}$ and $w_0$ is a free parameter.
The YBE reads
\be
R^{k_1,k_2}_{i_1,i_2}(u)R^{j_1,k_3}_{k_1,i_3}(u+v)
R^{j_2,j_3}_{k_2,k_3}(v) = R^{k_2,k_3}_{i_2,i_3}(v)
R^{k_1,j_3}_{i_1,k_3}(u+v)R^{j_1,j_2}_{k_1,k_2}(u)
\label{YBE-v}
\ee
for the vertex models and
\be
\sum_g\wt Wabgfu\wt {W}fgdev\wt {W}gbcd{v\!-\!u} \no \\
=\sum_g\wt {W}fage{v\!-\!u}\wt {W}abcgv\wt Wgcdeu \label{YBE-f} 
\ee
for the face models. The YBE provides a
sufficient condition for integrability of  models with
periodic boundary conditions. With open boundaries 
the integrability of exactly solvable models  is
guaranteed if in addition we have boundary Boltzmann weights $K$
which satisfy the boundary YBE
\be 
R^{k_1,k_2}_{i_1,i_2}(u-v)K^{l_1}_{k_1}(u;\xi)
 R^{j_1,l_2}_{l_1,k_2}(u+v)K^{j_2}_{l_2}(v;\xi)\nonumber \\
=K^{k_2}_{i_2}(v;\xi)R^{k_1,l_2}_{i_1,k_2}(u+v)K^{l_1}_{k_1}(u;\xi)
  R^{j_1,j_2}_{l_1,l_2}(u-v)
\label{BYBE-v}
\ee
for the vertex models and
\be
&&\hspace{-1cm}\sum_{f,g}{\wt Wgcba{u-v}}{\Km {K}gcf{u;\xi}}{\wt 
   Wdfga{u+v}}{\Km {K}dfe{v;\xi}} 
  \no\\
&& \hspace{-0.7cm}=\sum_{f,g} {\Km {K}bcf{v;\xi}}{\wt 
   Wgfba{u+v}}{\Km {K}gfe{u;\xi}
     }{\wt Wdega{u-v}} \label{BYBE-f}
\ee
for the face models. Generally we may have more arbitrary 
parameters than  $\xi$. The reflection equation 
(\ref{BYBE-v}) was first given by Cherednik\cit{Cherednik}. 
Then Sklyanin\cit{Sklyanin} applied it in the study of
the six-vertex model with open boundary conditions. 
The face analogue of the reflection equation (\ref{BYBE-f}) has
recently been written down  for $n=2$ 
only\cite{Kulish:95}$-$\cite{Zhou:95b}. For the higher rank face models 
we obtain (\ref{BYBE-f}) by converting the reflection equation 
(\ref{BYBE-v}) with the intertwiner given by Jimbo, Miwa and 
Okado\cit{JMO:87}.

To see  integrability let us define transfer matrices
$\T(u)$. Following Skylanin's formulation define the vertex
model transfer matrix for each $n\ge 2$ by
\be
\T(u)=\sum_{i,j,k,l=1}^nK^i_j(-n\lambda/2-u;\xi_+)
  \mbox{\boldmath $U$}_j^k(u)
  K_k^l(u;\xi_-)\tilde{\mbox{\boldmath $U$}}_l^i(u)  \label{T-v} 
\ee
where the $n\times n$ matrices $\mbox{\boldmath$U$}(u)$ and
$\tilde{\mbox{\boldmath $U$}}(u)$ in classical (auxiliary) space $c$ are
\be
\mbox{\boldmath$U$}(u)=
        R^{c,1}(u)R^{c,2}(u)\cdots,R^{c,N}(u) \label{U} \\
\tilde{\mbox{\boldmath $U$}}(u)=
   R^{N,c}(u)\cdots R^{2,c}(u)R^{1,c}(u).  \label{U1}
\ee
$\lambda$ is the crossing parameter. While for the face models, define the 
elements of the transfer matrix as  
\be && \hspace{-0.6cm}
\langle\mbox{\boldmath $a$}|\mbox{\boldmath $T$}(u)
   |\mbox{\boldmath $b$}\rangle
  =\sum_{\{c_0,\cdots,c_N\}} \sqrt{G_{c_0}^2\over G_{a_0}G_{b_0}}
  \Km {K}{c_0}{a_0}{b_0}{-n\lambda/2-u;\xi_+}  \nonumber\\ 
&&  \times\Km {K}{c_N}{a_N}{b_N}{u;\xi_-} 
 \prod_{k=0}^{N-1} \biggl[\wt {W}{b_k}{b_{k+1}}{c_{k+1}}{c_k}{u} 
 \wt {W}{c_{k+1}}{a_{k+1}}{a_k}{c_k}{u}\biggl],
     \label{openT}
\ee 
where  $\mbox{\boldmath $a$}=\{a_0,a_1,\cdots,a_N\}$ and 
$\mbox{\boldmath $b$}=\{b_0,b_1,\cdots,b_N\}$. 
$G_a$ is the crossing factor\cite{JMO:87}.
Using YBE and boundary YBE together with the inversion
relations\cit{RiTr:86,FHSY:95}
\be
R^{12}(u)R^{21}(-u)&=& \rho_1(u) \cdot {\rm id}\label{unity-v} \\
 \stackrel{\!\!t_1}{ R^{12}}(-n\lambda/2+u) 
 \stackrel{\!\!t_1}{R^{21}}(-n\lambda/2-u)&=& 
{\rho}_2(u) \cdot {\rm id}\;
\label{cross-v}
\ee
for the vertex models and\cit{JMO:87}
\be
\sum_{g} \wt Wabgdu\wt Wgbcd{-u} =\rho_1(u)\delta_{a,c}\label{unity-f} \\
\sum_{g} \wt Wdabg{-n\lambda/2-u} \wt
Wdgbc{-n\lambda/2+u} {G_{a}G_{g}\over
  G_{d}G_{b}}={\rho}_2(u)\delta_{a,c} \label{cross-f}
\ee
for the face models, we can prove the commutative relation
\be
\left[\mbox{\boldmath $T$}(u)\; ,
 \; \mbox{\boldmath $T$}(v)\;\right]=0\;. \label{comm}
\ee
This implies that the models are integrable.  
${\rho}_1(u),{\rho}_2(u)$ are 
scalar functions\cit{JMO:87} and the index $t_1$ denotes 
transposition in space $1$.  The  higher rank $A_{n-1}^{(1)}$ 
models  are special in the sense that the crossing 
symmetry does not exist. Nevertheless we still have  
integrability if the transfer matrices
are defined by (\ref{openT}). The formulation including the proof of 
(\ref{comm}) follows easily from Skylanin's work\cit{Sklyanin} 
(e.g. for the eight-vertex model\cit{VeGo:93,HoYu:93,InKo:94,Zhou:95b}, 
 for the  $\Z_n$-vertex models\cit{FHSY:95} and
for the $n=2$ face model\cit{Kulish:95}$-$\cite{Zhou:95b}). 

We have formulated the vertex and face models with 
open boundary conditions simultaneously.
Historically the face models came after the 
vertex models. A natural way to introduce the face models from
the vertex models is to use intertwiners\cite{Baxter:73,JMO:87}. 
These same intertwiners can also be used to convert the boundary 
Boltzmann weights\cite{Zhou:95b}. 
\subsection{Fusion hierarchies and functional relations} 
The fusion procedure\cit{KRS:81} has often been used to build new
exactly solvable models based on an elementary model for both
the periodic and open boundary cases.   We refer to some  
references\cit{DJKMO:88}$-$\cite{MeNe:92,BaRe:89,BaRe:90,Zhou:95b}
for the fusion of the preceding models. Here we will skip 
the details of fusion for models with open boundary
conditions. Instead we go straight to the 
functional relations for the fused models based on the elementary 
models.  

For periodic boundary conditions  the fused transfer matrices 
satisfy functional relations\cite{BaRe:89,BaRe:90}. The fusion levels
of the fused transfer matrices are labeled by two Young tableaux in 
vertical and horizontal directions. For unrestricted face and
vertex models, there is no  closure condition on the functional 
relations. Of particular interest here are the fused 
transfer matrices $\T^{(b,q)}(u):=\T^{(b,q)}_{(b^\prime,q^\prime)}(u)$ 
with rectangular Young tableaux of $b$($b^\prime$) rows and $q$($q^\prime$) 
columns. From  the known relations\cit{BaRe:90} the following $T$-system 
with $b^\prime,q^\prime$ fixed can be extracted\cit{KuNa:93}
\be 
&&\hspace{1cm}\T^{(b,q)}(u)\T^{(b,q)}(u-\lambda)\;=\; \nonumber\\
&&\T^{(b+1,q)}(u)\T^{(b-1,q)}(u-\lambda)+
 \T^{(b,q+1)}(u)\T^{(b,q-1)}(u-\lambda),  \label{func}
\ee
where $\T^{(b,q)}$ is an identity matrix if $b=0$ or $q=0$.
If $b=n$ the transfer matrices are reduced to  the diagonal matrices
with the common element being a model-dependent function
$f^q(u)$.  

For the above models with  open boundary conditions these functional
relations are still valid except that $f^q(u)$ picks up extra factors 
due to the boundaries.  

The functional equations can be easily converted into   
$y$-system equations in some sense\cite{KuNa:93,ZhPe:95}. Thus inserting
\be
\t^{(b,q)}(u):={\T^{(b,q+1)}(u)\T^{(b,q-1)}(u-\lambda)\over
\T^{(b+1,q)}(u)\T^{(b-1,q)}(u-\lambda)}     \label{def-t}
\ee 
into (\ref{func}) yields the thermodynamic Bethe ansatz-like  hierarchy
\be
\t^{(b,q)}(u)\t^{(b,q)}(u-\lambda)=
{[\1+\t^{(b,q+1)}(u)][\1+\t^{(b,q-1)}(u-\lambda)]\over
[\1+(\t^{(b+1,q)}(u))^{-1}][\1+(\t^{(b-1,q)}(u-\lambda))^{-1}]}
\label{TBA}\ee
with initial condition 
\be
\t^{(b,0)}(u)=0\;, \hspace*{0.5cm} b=1,2,\cdots,n-1\;.
\ee 

Here $\t^{(b,q)}(u)$ plays the role of finite-size 
corrections to the transfer matrices when the system size $N$ is large 
enough. Therefore  bulk and  surface free energies 
can be determined from the functional relations (\ref{func}) 
with the second term in the right hand side  omitted.  
This provides unitarity relations of the free energies 
for the special case $n=2$\cit{Zhou:95a,Zhou:95b}
(see eqn (3.3)\cit{Zhou:95a} for unitarity relations and 
section$~5.1$\cit{Zhou:95b} for a detailed
discussion). Making the proper assumption of 
analyticity the unitarity relations can be solved
by the standard technique employed by Baxter\cite{Baxter:82b}.
Thus some surface critical exponents can be extracted from
the leading singular part of surface free energies. 
This idea has successfully been applied to Baxter's 
eight-vertex model\cite{BaZh:95}, 
the ABF models\cite{ZhBa:95a}, the cyclic SOS models\cit{ZhBa:95b}
and the dilute SOS models\cite{BFZ:95}.
\subsection{$U_q[sl(2)]$--invariant case} In this part we focus on 
the $U_q[sl(2)]$-invariant vertex model. The same idea can
be applied to study $U_q[sl(n)]$-invariant vertex models.

For $U_q[sl(2)]$  we  go back to  results given 
by Kulish and Sklyanin\cite{KuSk:91}. 
The $U_q[sl(2)]$-invariant model is described
by the six-vertex model with  open boundary conditions,
which is  Baxter's eight-vertex model in the trigonometric limit
(critical limit). Therefore the  fused transfer matrices satisfy  
functional relations (\ref{func}) in the critical limit with $n=2$ and 
$\xi_\pm\to \infty$. Using the known Bethe ansatz results\cit{KuSk:91} 
it has been shown\cit{Zhou:95a} that the functional 
relations in the braid limit $u\to\pm{i}\infty$ are truncated at  
fusion level $h-1$ if the crossing parameter $\lambda=\pi/h$, 
where $h=4,5,\cdots$. Thus the functional relations can be used
to find the finite-size corrections to the fused transfer matrices
$T^{(p)}(u)$ with fusion level $p$ in both  the vertical and 
horizontal directions. In the large system size limit the fused transfer 
matrix eigenvalues are found to behave like\cit{Zhou:95a}
\be
\log T^{(p)}(u)=\!-\!2N f_b(u)\!-\!f_s(u)\!+\!{\pi\over 12N}
 \left(c\!-\!24\Delta_{1,\nu,s}\right)\sin(hu)
 \!+\!{ o\!}\left({1\over N}\right).
\label{finite}
\ee 
The central charge is given by 
\be
c&=&{3p\over p+2}-{6p\over h(h-p)}\;  \label{c} 
\ee
where $p=1,2,\cdots,h-2$ labels the fusion level.
The conformal weights are given by
\be
\Delta_{1,\nu,s}\;=\;{\left(h-(h-p)s\right)^2-p^2\over 4hp(h-p)} +
    {\nu(p-\nu)\over 2p(p+2)} \label{Kac}
\ee
with $\nu$ a unique integer determined by
\be 
\nu=s-1-p\lfloor{s-1\over p}\rfloor  \;. 
\ee
and $s=1,3,\cdots\le h-2$. Here $\lfloor{x}\rfloor$ is the largest
integer part less than or equal to $x$. The functions $f_b(u)$ and
$f_s(u)$ are respectively the bulk free energy and surface free 
energy of the critical model. They can be calculated by solving
the unitarity relation of the free energies as discussed in the 
last section. For the case of $p=1$ these conformal spectra 
coincide with  the known results\cite{SaBa:89}.
\subsection{Discussion} 
Here we have formulated the integrability
of Belavin's $\Z_n$ vertex models and  their relevant 
SOS models. It has been shown that  integrability for general $n>2$, 
strictly speaking, is guaranteed by the YBE and 
boundary YBE together with  
inversion relations ((\ref{unity-v})-(\ref{cross-v}) for vertex and
(\ref{unity-f})-(\ref{cross-f})
for face). The general formulation has also been applied to 
other models: $B^{(1)}_{n}$, $C^{(1)}_{n}$, $D^{(1)}_{n}$
and $A^{(2)}_{n}$\cite{BFKZ:95}. 

The functional relations of the fused transfer
matrices still obey the same fusion rules as in the periodic boundary
case. It has been shown  that the functional relations are useful
in finding both the bulk and surface free energies and the finite-size 
corrections to the transfer matrices of the models. 
\subsection*{Acknowledgments}
The author thanks Murray Batchelor and Vlad Fridkin 
for helpful comments. This work has been supported by the Australian 
Research Council and partially supported 
by the China Natural Science Foundation.
\def\AP#1#2#3{\newblock{\sl Ann. Phys.} {\bf #1} (#2) #3}
\def\CMP#1#2#3{\newblock{\sl Commun. Math. Phys.} {\bf #1} (#2) #3}
\def\LMP#1#2#3{\newblock{\sl Lett. Math. Phys.} {\bf #1} (#2) #3}
\def\JSP#1#2#3{\newblock{\sl J. Stat. Phys.} {\bf #1} (#2) #3}
\def\JPA#1#2#3{\newblock{\sl J. Phys.} {\bf A#1} (#2) #3}
\def\JMP#1#2#3{\newblock{\sl J. Math. Phys.} {\bf #1} (#2) #3}
\def\IJMP#1#2#3{\newblock{\sl Int. J. Mod. Phys.} {\bf #1} (#2) #3}
\def\MPLA#1#2#3{\newblock{\sl Mod. Phys. Lett.} {\bf A#1} (#2) #3}
\def\NPB#1#2#3{\newblock{\sl Nucl. Phys.} {\bf B#1} (#2) #3}
\def\PLA#1#2#3{\newblock{\sl Phys. Lett.} {\bf A#1} (#2) #3}
\def\PLB#1#2#3{\newblock{\sl Phys. Lett.} {\bf B#1} (#2) #3}
\def\PRL#1#2#3{\newblock{\sl Phys. Rev. Lett.} {\bf#1} (#2) #3}
\def\PR#1#2#3{\newblock{\sl Phys. Rev.} {\bf#1} (#2) #3}
\def\PTP#1#2#3{\newblock{\sl Prog. Theor. Phys. } {\bf#1} (#2) #3}
\def\LMP#1#2#3{\newblock{\sl Lett. Math. Phys.} {\bf#1} (#2) #3}
\def\PA#1#2#3{\newblock{\sl Physica A } {\bf#1} (#2) #3}
\def\SPJ#1#2#3{\newblock{\sl Sov. Phys. JETP. } {\bf#1} (#2) #3}


\begin{thebibliography}{99}
\bibitem{Yang:67} C. N. Yang, \PRL {19}{1967}{1312.}
\bibitem{Baxter:82} R. J. Baxter, ``Exactly Solved Models in Statistical
Mechanics",   Academic Press, London, 1982.
\bibitem{Cherednik}I. Cherednik, 
  {\sl Theor. Math. Phys.} {\bf 61}(1984) 977.
\bibitem{Sklyanin}E. K. Sklyanin, \JPA {21}{1988}{2375}.
\bibitem{GhZa:94}S. Ghoshal and A. Zamolodchikov,
           \IJMP {A 21}{1994}{3841}.
\bibitem{VeGo:93}H. J. de Vega and A. Gonz\'alez Ruiz, 
  \JPA {26}{1993}{L519}; \NPB {417}{1994}{553}; \PLB {332}{1994}{123}.
\bibitem{HoYu:93}B. Y. Hou and R. H. Yue,    \PLA {183}{1993}{169};
   B. Y. Hou,  K. J. Shi, H. Fan and Z. X. Yang,
      {\sl Commun. Theor. Phys.} {\bf 23} (1995) 163.
\bibitem{InKo:94}T. Inami and H. Konno, \JPA {27}{1994}{L913.}
\bibitem{Murray} M. T. Batchelor, {\sl Reflection equations and surface 
  critical phenomena}, to appear in the proceedings of statistical 
  mechanics models,
  Yang-Baxter equations and related topics, Tianjin meeting  Aug 1995.
\bibitem{FoKa:93}A. Foerster and M. Karowski, \NPB {408}{1993}{512}.
\bibitem{Ruiz94}A. Gonz\'alez Ruiz, \NPB {424}{1994}{468}.
\bibitem{MNR:90}L. Mezincescu, R. I. Nepomechie and V. Rittenberg, 
        \PLB {147}{1990}{70}.
\bibitem{AMN94}S. Artz, L. Mezincescu and R. I. Nepomechie, 
  \JPA {28}{1995}{5131}.
\bibitem{JKKKM}M. Jimbo, R. Kedem, T. Kojima, H. Konno 
  and T. Miwa, \NPB {441}{1995}{437.} 
\bibitem{talk}P. A.  Pearce, 
  {\sl Exact solution of lattice spin models with fixed boundary 
     conditions}, talk at the Tianjin meeting  Aug 1995.
\bibitem{MeNe:95} L. Mezincescu and R. I. Nepomechie, 
  ``Lectures on Integrable  Quantum Spin Chains,'' 
  in {\em New Developments of Integrable Systems and Long Ranged 
  Interaction Models}, ed. by M.-L. Ge and Y.-S. Wu,
  World Scientific Publishing Co, 96 - 142 (1995); \NPB {372}{1992}{579}.
\bibitem{YuBa:95} C. M. Yung and M. T. Batchelor, \NPB {435}{1995}{430}.
\bibitem{Kulish:95}P. P. Kulish {\em ``Yang-Baxter and reflection 
  equations   in integrable models'', }   hep-th/9507070.  
\bibitem{BPO:95}R. E. Behrend, P. A. Pearce and D. L. O'Brien 
 {\em ``Interaction-Round-a -Face   models with fixed boundary 
  conditions: the ABF fusion hierarchy'', }  hep-th/9507118. 
\bibitem{AhKo:95}  C. Ahn and W. M. Koo,
  {\em  ``Boundary Yang-Baxter equation in the RSOS representation",}
    hep-th/9508080.
\bibitem{Zhou:95b}Y. K. Zhou, \NPB {458}{1996}{} in press. 
\bibitem{eight} R. J. Baxter, 
  \AP {70}{1972}{193}.

\bibitem{Belavin} A. A. Belavin, \NPB {180}{1981}{189}. 
\bibitem{Baxter:73}R. J. Baxter, \AP {76}{1973}{1}; 25; 48. 
\bibitem{JMO:87} M. Jimbo, T. Miwa and M. Okado, 
   \LMP {14}{1987}{123}; \newline   \CMP {116}{1988}{507.}
\bibitem{Baxter:82b}R. J. Baxter,\JSP {28}{1982}{1} and references therein.
\bibitem{BaRe:89} V.~V.~Bazhanov and N.~Yu.~Reshetikhin,
          \newblock \IJMP {B4}{1989}{115} and references therein.
\bibitem{BaRe:90} V. V. Bazhanov and N. Yu Reshetikhin, \JPA
     {23}{1990}{1477} and references therein.
\bibitem{HYZ:89} B. Y. Hou, M. L. Yan and Y. K. Zhou, \NPB
                   {324}{1989}{715}.
\bibitem{RiTr:86} M. P. Richey  and C. A. Tracy   \JSP {42}{1986}{311}. 
\bibitem{FHSY:95} I. Cherednik,  {\sl Theor. Math. Phys.} {\bf 17} (1983) 77;
   H. Fan, B. Y. Hou, K. J. Shi and Z. X. Yang,
   \PLB {200}{1995}{109}. 
\bibitem{KRS:81} P.~P.~Kulish, N.~Yu.~Reshetikhin and E.~K.~Sklyanin,
     \LMP {5}{1981}{393}.
\bibitem{DJKMO:88} E. Date, M. Jimbo, A. Kuniba, T. Miwa and M. Okado,
           Adv. Stud. Pure Math.,{\bf 16} (1988) 17.
\bibitem{JKMO:88} M. Jimbo, A. Kuniba, T. Miwa and M. Okado, 
     \CMP {119}{1988}{543.}
\bibitem{ZhHo:89} Y. K. Zhou and B. Y. Hou, \JPA {22}{1989}{5089}.
\bibitem{MeNe:92}L. Mezincescu and R. I. Nepomechie, 
 \JPA {25}{1992}{2533}.
\bibitem{KuNa:93} A.Kuniba and T.Nakanishi, 
  {\sl Int. J. Mod. Phys. (Proc. Suppl.)} {\bf A3}(1993) 419.
\bibitem{ZhPe:95}Y. K. Zhou and P. A. Pearce, \NPB {446}{1995}{485}.
\bibitem{KuSk:91}P. P. Kulish and E. K. Sklyanin, \JPA {24}{1991}{L435}.
\bibitem{Zhou:95a}Y. K. Zhou, \NPB {453}{1995}{619}.
\bibitem{BaZh:95}M. T. Batchelor and Y. K. Zhou, {\sl Phys. Rev. Lett.}
          in press.
\bibitem{ZhBa:95a}Y. K. Zhou and M. T. Batchelor, 
              {\sl Surface Critical Phenomena in 
              Interaction-Round-a-Face Models,}  cond-mat/9511008.
\bibitem{ZhBa:95b} Y. K. Zhou and M. T. Batchelor, 
       {\sl Exact solution and surface critical behaviour
       of open cyclic SOS lattice models,}  preprint (1995). 
\bibitem{BFZ:95} M. T. Batchelor, V. Fridkin and Y. K. Zhou, 
     {\sl An Ising model in a magnetic field
      with a boundary,}   cond-mat/9511090. 
\bibitem{SaBa:89}H. Saleur and Bauer, \NPB {320}{1989}{591}.
\bibitem{BFKZ:95} M. T. Batchelor, V. Fridkin, A. Kuniba 
    and Y. K. Zhou, {\sl Solutions of reflection
    equation for face and vertex models associated with $A^{(1)}_{n}$, 
    $B^{(1)}_{n}$, $C^{(1)}_{n}$, 
    $D^{(1)}_{n}$ and  $A^{(2)}_{n}$,} preprint (1995). 
\end{thebibliography}
\end{document}